\documentclass[pra,onecolumn,superscriptaddress]{revtex4-2}
\usepackage{natbib}
\usepackage{amssymb,amsmath}
\usepackage{graphicx}
\usepackage{verbatim}
\usepackage{multirow}
\usepackage{graphicx}
\usepackage{bm}
\usepackage{mathtools}
\usepackage{subcaption}
\usepackage{ragged2e}

\usepackage{siunitx}
\usepackage[ruled]{algorithm2e}
\usepackage{multirow}
\usepackage{tikz}
\usetikzlibrary{arrows.meta, positioning}
\usepackage{adjustbox}
\usepackage{colortbl}
\usepackage{ulem}

\def\bR{{\bf R}}
\def\br{{\bf r}}
\def\bL{{\bf L}}
\def\bl{{\bf l}}
\newcommand{\bN}{\mbox{\boldmath$\nabla$}}

\begin{document}

\title{Quantum Computing Approach to  Atomic and Molecular Three-Body Systems}

\author{Mohammad Haidar}
\affiliation{OpenVQA Hub and WYW, 60 rue François Ier, F-75008 Paris, France}
\author{Hugo D. Nogueira}
\affiliation{Laboratoire Kastler Brossel, Sorbonne Universit\'{e}, CNRS, ENS-Universit\'e PSL, Coll\`{e}ge de France, 4 place Jussieu, F-75005 Paris,France}
\author{Jean-Philippe Karr}
\affiliation{Laboratoire Kastler Brossel, Sorbonne Universit\'{e}, CNRS, ENS-Universit\'e PSL, Coll\`{e}ge de France, 4 place Jussieu, F-75005 Paris,France}
\affiliation{Universit\'{e} Evry Paris-Saclay, Boulevard Fran\c{c}ois Mitterrand, F-91000 Evry, France}


\begin{abstract}
We present high-precision quantum computing simulations of three-body atoms (He, H$^-$) and molecules (H$_2^+$, HD$^+$), the latter being studied beyond the Born–Oppenheimer approximation. The Non-Iterative Disentangled Unitary Coupled Cluster Variational Quantum Eigensolver (NI-DUCC-VQE) [M. Haidar \textit{et al.}, Quantum Sci. Technol. \textbf{10}, 025031 (2025)] is used. By combining a first-quantized Hamiltonian with a Minimal Complete Pool (MCP) of Lie-algebraic excitations, we construct a compact ansatz with a gradient-independent construction, avoiding costly gradient evaluations and  yielding efficient computational scaling with both basis size and electron number. It avoids barren plateaus and enables rapid convergence, achieving energy errors as low as $10^{-11}$ a.u. with state fidelities only limited by arithmetic precision in only a few thousand function evaluations in all four systems. These results make three-body atoms and molecules excellent candidates for benchmarking and testing on current Noisy Intermediate-Scale Quantum (NISQ) devices. Further, our approach can be extended to more complex systems with larger basis sets, taking advantage of the efficient scaling of qubit requirements to study electronic correlations and non-adiabatic effects with high precision.  We also demonstrate the applicability of NI-DUCC-VQE for simulating higher-order effects such as relativistic corrections and hyperfine interactions.
\end{abstract}

\maketitle

\section{Introduction}

The quantum three-body problem has long been a computational challenge for physicists. In the absence of general analytical solutions, one has to rely on numerical methods for precise results. Over decades, classical computing approaches — particularly the variational method — have achieved exceptional accuracy on systems such as the helium atom and H$^-$ ion~\cite{Korobov2000,Schwartz2006b,Schwartz2006,Nakashima2007,Frolov2015,Aznabaev2018,Drake2023,Petrimoulx2025} as well as hydrogen molecular ions (HMI)~\cite{Korobov2000,Cassar2004,Hijikata2009,Ning2014}. Whereas high-precision calculations in He and HMI are of high relevance for fundamental physics tests and determination of fundamental constants (see e.g.~\cite{Pachucki2017,Schiller2022} for reviews), numerical results have now surpassed any realistic requirement. For example, the ground-state energies of the helium atom and H$_2^+$ molecule have been obtained with 45 and 33 significant digits, respectively~\cite{Schwartz2006,Ning2014}. The H$^-$ ion is also among the systems that have been studied with very high precision~\cite{Frolov2015,Aznabaev2018,Petrimoulx2025}, as the sensitivity of its only bound state to electron correlation makes it a critical test for numerical methods.

Meanwhile, recent advances in quantum computing methods~\cite{aspuru2005simulated,reiher2017elucidating,bassman2021simulating, kitaev1995quantum, peruzzo2014variational, mcardle2020quantum, cao2019quantum, bauer2020quantum, yeter2021benchmarking} offer promising perspectives to describe electron correlation effects with high precision and treat molecular structure beyond the Born-Oppenheimer approximation. In this context, the above-mentioned well-understood three-body systems serve as excellent reference tests for evaluating quantum algorithms and assessing the performance of near-term Noisy Intermediate-Scale Quantum (NISQ) devices currently being explored across academia and industry~\cite{whitfield2011simulation, peruzzo2014variational,preskill2018quantum, elfving2020will, bharti2022noisy}. The Variational Quantum Eigensolver (VQE)~\cite{peruzzo2014variational,google2020hartree, kandala2017hardware, robledo2024chemistry,hempel2018quantum,ollitrault2024estimation,haidar2022open} stands out among quantum algorithms for estimating ground-state energies. Variants like UCC-VQE~\cite{romero2018strategies,sokolov2020quantum,xia2020qubit,anand2021quantum,fedorov2022unitary,haidar2023extension}, ADAPT-VQE~\cite{grimsley2019adaptive,tang2021qubit,liu2021efficient,yordanov2021qubit,anastasiou2024tetris,shkolnikov2023avoiding,feniou2023overlap,traore2024shortcut,ramoa2025reducing} and other recent promising approaches such as in~\cite{ryabinkin2018qubit,ryabinkin2018constrained,ryabinkin2020iterative,lang2020unitary,lang2023growth,burton2024accurate,mondal2023development,halder2024noise,patra2024toward,halder2025construction,Patra2025VQE} aim to balance expressiveness
and hardware efficiency. In this work, we focus on the Non-Iterative Disentangled Unitary Coupled Cluster VQE (NI-DUCC-VQE), introduced by M.~Haidar et al.~\cite{haidar2025non}. This algorithm removes the need for gradient evaluations, simplifies circuit design, and still achieves accurate results in molecular simulations. 
We apply NI-DUCC-VQE to a few representative three-body systems - HMI (H$_2^+$, HD$^+$), which we study without relying on the Born-Oppenheimer approximation, as well as the H$^-$ and helium atoms. We compare our quantum results against highly accurate classical benchmarks from variational methods. This comparison demonstrates the precision and promise of NI-DUCC-VQE for quantum simulation of  more complex systems. While most quantum chemistry algorithms operate in the second quantization formalism~\cite{jordan1993paulische,ortiz2001quantum}, we instead use a first-quantized Hamiltonian~\cite{abrams1997simulation}. This approach offers major advantages in qubit efficiency, with required qubits scaling as $n_e \cdot \log_2 N$, where $n_e$ is the number of electrons and $N$ is the number of basis functions~\cite{Volkmann2024,georges2025quantum,per2025chemically}. This favorable scaling enables the use of larger basis sets while keeping resource requirements within the range of NISQ devices. In fact, the promise of the first-quantization formalism has already been tested experimentally, in a recent helium calculation using a Quantinuum ion-trap quantum computer \cite{per2025chemically}. We present the first implementation of NI-DUCC-VQE with first quantization. This combination eliminates gradients, reduces circuit depth, and benefits from compact wavefunction encoding. Together, these features establish a scalable and hardware-efficient approach for simulating three-body quantum systems with high accuracy.

\section{Nonrelativistic three-body Hamiltonian and exponential basis functions}
\label{sec:threebody}

We consider a general three-body atom or molecule. The mass and charge of particle $i$ are represented by $m_i$ and $Z_i$, respectively, and its position in the laboratory frame is described by the variable $\bR_i$. The nonrelativistic Hamiltonian in the center-of-mass frame is given by the expression \cite{Drake2023}:
\begin{equation}
\label{Hamiltoniantotal}  
H = -\frac{1}{2m_{12}}\bN^2_{\br_1}-\frac{1}{2m_{13}}\bN^2_{\br_2}-\frac{1}{m_1}\bN_{\br_1}\cdot\bN_{\br_2} + \frac{Z_1Z_2}{r_1} + \frac{Z_1Z_3}{r_2} + \frac{Z_2Z_3}{R},
\end{equation}
where $m_{ij} = m_i m_j/(m_i + m_j)$, $\br_1 = \bR_2 - \bR_1$, $\br_2 = \bR_3 - \bR_1$, and $\bR = \bR_3 - \bR_2$. Note that all the quantities are expressed in atomic units, with masses in units of the electron's mass and distances in units of Bohr's radius, 
making them dimensionless. The solutions of the three-body Schr\"odinger equation can be classified according to the values of the quantum numbers $L$ (orbital angular momentum), $M$ (its projection on the internuclear axis), and $\Pi$ (parity). The total angular momentum can be written as \[
\bL = \bl_1 + \bl_2, \quad \bl_i = -i \br_i \times \nabla_{\br_i}.\]
The wavefunction of a state with given values of $L$, $M$ and $\Pi$ can be separated into angular and radial parts as follows~\cite{Schwartz1961}:
\begin{equation}
\label{Psi_total}
\psi_{LM}^\Pi = \sum_{l_1 + l_2 = L_\Pi} r_1^{l_1} r_2^{l_2} \left\{ Y_{l_1}(\hat{\br}_1) \otimes Y_{l_2}(\hat{\br}_2) \right\}_{LM} F_{l_1 l_2}^{L\Pi}(r_1, r_2, R).
\end{equation}

The coupled product of two spherical harmonics in the above expression is called a bipolar spherical harmonic. These functions form a basis of eigenfunctions of $\bL^2$, $L_z$, and $\Pi$, with respective eigenvalues $L(L+1)$, $M$, and $(-1)^{l_1 + l_2}$. The value of $L_\Pi$ depends on parity: $L_{\Pi}=L$ if $\Pi = (-1)^L$, and $L_{\Pi}=L+1$ if $\Pi = -(-1)^L$.
In the present work, we focus on the ground electronic state, where $L=M=0$ and $\Pi = (-1)^L$. The three-body wavefunction then reduces to the radial wavefunction $F_{00}^{01} (r_1,r_2,R) \equiv F(r_1,r_2,R)$ in Eq.~(\ref{Psi_total}), which can be expanded in a suitable basis. One particularly effective choice is the expansion using exponential functions~\cite{Korobov2000,Schwartz2006b,Frolov2015,Aznabaev2018}, which takes the form: 
\begin{equation}
\label{basisfunction}
F (r_1,r_2,R) = \sum_{i=1}^{N} \left[ C_i {\rm Re} (e^{-\alpha_i r_1 - \beta_i r_2 - \gamma_i R}) + D_i {\rm Im} (e^{-\alpha_i r_1 - \beta_i r_2 - \gamma_i R}) \right],
\end{equation}
where $\alpha_i, \beta_i, \gamma_i$ are complex exponents. In molecular systems, labeling the electron as particle 1, the $R$ variable corresponds to the internuclear distance. Then, the choice of complex $\gamma_i$ allows for capturing the oscillatory behavior of vibrational states in molecular systems. While $\alpha_i$ and $\beta_i$ can be real, adding a (small) imaginary part increases flexibility and slighlty improves convergence. In atomic systems, labeling the nucleus as particle 1, $R$ corresponds to the interelectronic distance. The presence of the $-\gamma_i R$ term in the expansion~(\ref{basisfunction}) signals the use of explicitly correlated basis functions. Optimizing all $3N$ nonlinear parameters individually would be computationally prohibitive. Instead, they are generated pseudo-randomly within predefined intervals, significantly reducing the number of variational parameters. It should be noted that this approach prevents the use of gradient-based optimization techniques, as the Hamiltonian matrix elements depend only on the interval bounds rather than on the explicit parameters.

For homonuclear molecules such as H$_2^+$, the wavefunction exhibits an additional symmetry: reflection through a plane perpendicular to the internuclear axis and passing through the midpoint of the nuclei. Consequently, the wavefunction must be either symmetric or antisymmetric under the exchange $r_1 \leftrightarrow r_2$. In the ground state, the wavefunction is symmetric. Similarly, in two-electron atomic systems,  the wavefunctions exhibit symmetry properties under electron exchange, distinguishing singlet (symmetric) and triplet (antisymmetric) states; again, the ground state is symmetric. In those cases, we exploit these symmetries by using symmetrized basis functions (adding in Eq.~(\ref{basisfunction}) a term where the roles of $r_1$ and $r_2$ are exchanged), improving computational efficiency and accuracy. We also consider in the following the heteronuclear molecule HD$^+$, allowing to study the effect of breaking this symmetry.

In the next section, we describe how to transform the non-relativistic three-body problem into the language of quantum information. After presenting the NI-DUCC ansatz, we map the three-body Hamiltonian in its first-quantized form to a qubit representation for implementation within the NI-DUCC-VQE algorithm.

\section{NI-DUCC-VQE} \label{sec:NI-DUCC}

The NI-DUCC-VQE method~\cite{haidar2025non} combines the advantages of the NI-DUCC wavefunction and the Variational Quantum Eigensolver (VQE).  NI-DUCC provides a compact and structured ansatz, while VQE optimizes its parameters to minimize the energy. 

\subsection{NI-DUCC wavefunction} \label{NI-DUCCWavefunction}

The disentangled wavefunction is represented as a product of unitary transformations that act on a reference state $|\psi_0\rangle$~\cite{evangelista2019exact}:
\[
|\Psi\rangle = \prod_{l} e^{\theta_l \hat{T}_l} |\psi_0\rangle,
\]
 where each \(\hat{T}_l\) is an excitation operator with parameter \(\theta_l\). Unlike traditional Coupled-Cluster (CC) methods~\cite{anand2021quantum}, where \( \hat{T_l} \) is built from fermionic excitation operators, the non-iterative disentangled Unitary Coupled-Cluster (NI-DUCC) ansatz adopts a different strategy. The CNOT staircase method decomposes the exponential of an  operator into single- and two-qubit gates, and,  when  applied to fermionic operators,  is not hardware-efficient because it generates many CNOT gates~\cite{mcardle2020quantum,whitfield2011simulation,hempel2018quantum}. Since CNOT gates are the main source of noise in current quantum hardware, this is not desirable for NISQ devices~\cite{preskill2018quantum}. This motivates the construction of a new operator pool formed by Pauli strings that involve fewer CNOTs, thereby improving compatibility with near-term NISQ hardware. In NI-DUCC~\cite{haidar2025non}, the cluster operator \( \hat{T} \) is constructed using Pauli strings formed through the use of a minimal complete pool (MCP) of size \( 2n-2 \), where $n$ is the number of qubits. The MCP ensures both the preservation of a Lie algebraic structure and a reduction in computational cost in particular through the smaller number of CNOT gates. The existence of such MCP has been rigorously proven in Ref.~\cite{shkolnikov2023avoiding}. In particular, one family of MCPs introduced in Ref.~\cite{tang2021qubit}, denoted 
\( \{V_j\}_n \) for \( j = 1, \ldots, 2n-2 \), is defined recursively as  
\begin{equation}
\{V_j\}_n = \{ Z_n \{V_k\}_{n-1},\; iY_n,\; iY_{n-1} \},
\end{equation}
where the set contains single-qubit \( iY \) rotations as well as conditional two-qubit 
\( iZ_{k+1}Y_k \) rotations that act locally on neighboring qubits. 
Here \( I \) denotes the \(2 \times 2\) identity matrix, and 
\( X \equiv \sigma_x \), \( Y \equiv \sigma_y \), \( Z \equiv \sigma_z \). 
Moreover, these operators form a set that is closed under commutation, meaning that for any 
\( A, B \in \{V_j\}_n \), the commutator \([A,B] = AB - BA\) is again contained in the span 
of \(\{V_j\}_n\). In other words, they generate a Lie algebra and are sufficient to rotate any real-valued state within the Hilbert space (see Appendix A in Ref.~\cite{shkolnikov2023avoiding}). The corresponding unitary transformation reads 
\begin{equation}
\hat{U}(\vec{\theta}) = \prod_{l=1}^{2n-2} e^{i \theta_l \hat{P}_l},
\end{equation}
where each \( \hat{P}_l \) is a selected Pauli operator from the MCP \( \{V_j\}_n \) as defined in Ref.~\cite{shkolnikov2023avoiding}. The adoption of an MCP guarantees closure under commutation and thus a well-defined 
Lie-algebraic structure. This provides a solid foundation for the systematic construction 
of the wavefunction ansatz and accelerates convergence to the target state, thereby 
reducing optimization costs in practical applications.
To enhance the expressivity of the ansatz while maintaining circuit efficiency, NI-DUCC employs a layer-wise construction comprising \( k \) layers:
\begin{equation}
\label{disentangled_comp}
|\Psi(\vec{\theta})\rangle = \left[ \prod_{m=1}^{k} \left( \prod_{l=1}^{2n-2} e^{i \theta_l \hat{P}_l} \right) \right] |\psi_{0}\rangle.
\end{equation}
This hierarchical strategy allows for systematic improvement in accuracy while keeping the number of gates and circuit depth manageable, thus ensuring compatibility with noisy quantum hardware. The Variational Quantum Eigensolver (VQE) algorithm is then employed to optimize the variational parameters \( \theta_l \) that characterize the NI-DUCC ansatz. 

\subsection{First-quantized Hamiltonian in NI-DUCC-VQE} \label{first_quantizedsec}

We now define the three-body Hamiltonian in a qubit-compatible form, enabling  efficient implementation of the above-described approach.
\\
As explained in Sec.~\ref{sec:threebody}, we expand the ground-state wavefunction (which is reduced to a radial wavefunction $F(r_1,r_2,R$)) in terms of the basis functions $\phi_i$,
\begin{equation}
\label{Totalwavefunction1}
F (r_1,r_2,R) = \sum_{i=1}^N c_i \, \phi_i (r_1,r_2,R) \,,
\end{equation}
where 
\begin{equation}
\label{basisfunc}
\phi_i (r_1,r_2,R) = \left[ \mathrm{Re \; or \; Im} \right] \left(e^{-\alpha_i r_1 - \beta_i r_2 - \gamma_i R} \right) \pm (r_1 \leftrightarrow r_2) \,.
\end{equation}
In the above expression, the second term is added for symmetric systems, i.e. the homonuclear molecule H$_2^+$ and two-electron atoms, with a positive sign in the case of the ground state. Application of the variational principle leads to the generalized eigenvalue problem~\cite{Drake2023}
\begin{equation} \label{eq-generalized eig}
\textbf{H} \textbf{c} = E \textbf{O} \textbf{c},
\end{equation}
with $H_{ij} = \langle \phi_i | H | \phi_j \rangle$ and $O_{ij} = \langle \phi_i | \phi_j \rangle$. These matrix elements are calculated analytically~\cite{Drake2023}. The presence of the overlap matrix $\mathbf{O}$ on the right-hand side is due to the fact that the basis functions~(\ref{basisfunc}) are not orthogonal to each other. As it is easier to work with an orthonormal basis set, where $\mathbf{O}$ simply becomes the identity matrix, we follow \cite{Volkmann2024} and apply an orthonormalization procedure. There are two kinds of procedures, the sequential ones, in which each eigenvector is orthogonalized with respect to the previously orthogonalized one, such as the Gram-Schmidt method (used in \cite{Volkmann2024}), and those in which all eigenvectors are simultaneously orthogonalized \cite{Jiao2015}. We decided to use the canonical orthogonalization method proposed by \cite{Lowdin1970}, which is of the second kind and is described in Appendix~\ref{app_co}. Besides making the calculation simpler, using an orthonormalized basis set can avoid the appearance of numerical issues due to ill conditioned matrices \cite{Volkmann2024, Jiao2015}. 

After orthonormalization, the first-quantized Hamiltonian is represented as~\cite{Volkmann2024}
\begin{equation}
\label{first_quantized}
H = \sum_{i,j = 1}^N H_{ij} |\phi_i \rangle \langle \phi_j | \,.
\end{equation}
where we have kept our previous notations $\phi_i$, $H_{ij}$, which now apply to the orthonormalized basis functions and transformed Hamiltonian matrix.

Finally, the qubit representation of the Hamiltonian operator (\ref{first_quantized}) can be obtained by employing binary encoding~\cite{abrams1997simulation,Tilly2022}. For instance, the qubit state \( |00 \ldots 00\rangle \) represents \( |\phi_1\rangle \), \( |10 \ldots 00\rangle \) represents \( |\phi_2\rangle \), and so forth. An operator such as \( |\phi_1\rangle \langle \phi_2| \) is then expressed as $
|\phi_1\rangle \langle \phi_2 | = |00 \ldots 00\rangle \langle 10 \ldots 00 | = \left(|0\rangle \langle 1|\right) \otimes \left(|0\rangle \langle 0|\right) \otimes \ldots \otimes \left(|0\rangle \langle 0| \right)\,.$
 This shows that all operators can be rewritten into a tensor product of four types of single-qubit projectors: \( |0\rangle \langle 0| \), \( |0\rangle \langle 1| \), \( |1\rangle \langle 0| \), and \( |1\rangle \langle 1| \), which are mapped to Pauli operators as follows:
$|0\rangle \langle 0| = \frac{1}{2} (I + Z), \quad |0\rangle \langle 1| = \frac{1}{2} (X + iY), \quad |1\rangle \langle 0| = \frac{1}{2} (X - iY), \quad |1\rangle \langle 1| = \frac{1}{2} (I - Z)$.
As a result, the Hamiltonian is represented as a sum of  Pauli strings, each of length \( \mathcal{O}(n) \) (with a qubit number $n=\log_2(N)$), enabling efficient measurement on quantum devices~\cite{mcardle2020quantum,haidar2022open}.

The variational quantum eigensolver (VQE) is a hybrid quantum--classical approach. 
It relies on repeated measurements of the Hamiltonian expectation value to guide the 
classical optimization, driving convergence toward the ground-state energy of the system. 
Once optimized, the variational parameters define an approximate ground-state 
wavefunction with high fidelity. Within the NI-DUCC formalism, this provides an 
efficient framework for studying correlated quantum systems. The high performance of the NI-DUCC VQE has been demonstrated in~\cite{haidar2025non} for the LiH, H$_6$ and BeH$_2$ molecules using a second-quantization approach. In what follows, we describe its implementation for three-body systems in the above-described first-quantization formalism and present our numerical results.

\section{Computational details}
This study involves four quantum systems: H$_2^+$, HD$^+$, He, and H$^-$. We focused on the ground ro-vibrational states ($1s\sigma$, $L = 0$, $v = 0$) of H$_2^+$, HD$^+$ and the ground electronic states ($1^1S$) of He and H$^-$. For each system, simulations were performed using 7 and 8 qubits, corresponding to 128 and 256 basis functions, respectively. The corresponding exponential basis sets, as described in Sec.~\ref{sec:threebody} and Eq.~(\ref{basisfunc}), are generated by sampling the real and imaginary parts of the exponents $\alpha_i$, $\beta_i$, and $\gamma_i$  within several bounded intervals.

Optimizing these bounds is essential to ensure high accuracy 
since the Hamiltonian matrix elements do not explicitly depend on the interval bounds, gradient-based optimization methods are not applicable. Therefore, we used the gradient-free BOBYQA algorithm~\cite{Powell2009}, starting from physically motivated initial bounds. The Tables of Appendix~\ref{referencesclassical} summarize the results for all the systems and basis sizes considered in this work.

The Hamiltonian and overlap matrix elements were analytically calculated using recurrence relations~\cite{Drake2023,Haidar2021} and evaluated with quadruple precision using Fortran codes. Löwdin symmetric orthonormalization~\cite{Lowdin1956} was then applied to eliminate the overlap matrix and remedy numerical near-linear dependency
problems that potentially arise for sufficiently large bases at limited numerical precision. This provides us with highly precise values of the matrix elements $H_{ij}$ required for representing the first-quantized Hamiltonian~(\ref{first_quantized}), which are subsequently stored as double-precision numbers, all the digits being significant.

Following the construction of the Hamiltonian, the NI-DUCC wavefunctions [Eq.~(\ref{disentangled_comp})] and associated energy levels were computed using Python-based development codes. The action of exponentiated Pauli operators on the reference state $|0\rangle$ was evaluated using the $\mathrm{expm\_multiply}$ function of SciPy~\cite{2020SciPy-NMeth}. The MCP of Pauli excitation operators were constructed and validated using custom Python routines. The VQE parameters were optimized using the BFGS algorithm of \texttt{scipy.optimize}~\cite{2020SciPy-NMeth}, using analytically computed gradients and a convergence criterion based on a gradient norm of $10^{-10}$ a.u. To benchmark performance, NI-DUCC-VQE was compared with Qubit-ADAPT-VQE~\cite{tang2021qubit}. 
While both methods employ Pauli operators, Qubit-ADAPT-VQE constructs its ansatz adaptively 
by choosing operators according to the energy gradients, whereas NI-DUCC-VQE avoids such 
gradient evaluations.

\section{Numerical results}

\subsection{NI-DUCC-VQE: energy convergence behavior}

\begin{figure}[htp!]
    \centering
    \begin{subfigure}[b]{\textwidth}
        \centering
        \begin{tabular}{ccccc}
            \hline
            $N$ & H$_2^+$ & HD$^+$ & He & H$^-$ \\
            \hline
            $E_{N=128}$ & $-0.597\,139\,058\,413$ & $-0.597\,897\,235\,547$ & $-2.903\,724\,376\,970$ & $-0.527\,751\,016\,439$ \\
            $E_{\rm ref}$ & $-0.597\,139\,063\,080$ & $-0.597\,897\,968\,103$ & $-2.903\,724\,377\,034$ & $-0.527\,751\,016\,544$ \\
            \hline
        \end{tabular}
        \caption{Ground-state energy values for the four systems. First line: results of the classical variational method  with $N = 128$. Second line: reference values obtained with a much larger value of $N$, where all printed digits are exact. Energies are in atomic units.}
        \label{fig:subfig-b}
    \end{subfigure}
    \begin{subfigure}[b]{0.495\textwidth}
        \centering
        \includegraphics[width=\textwidth]{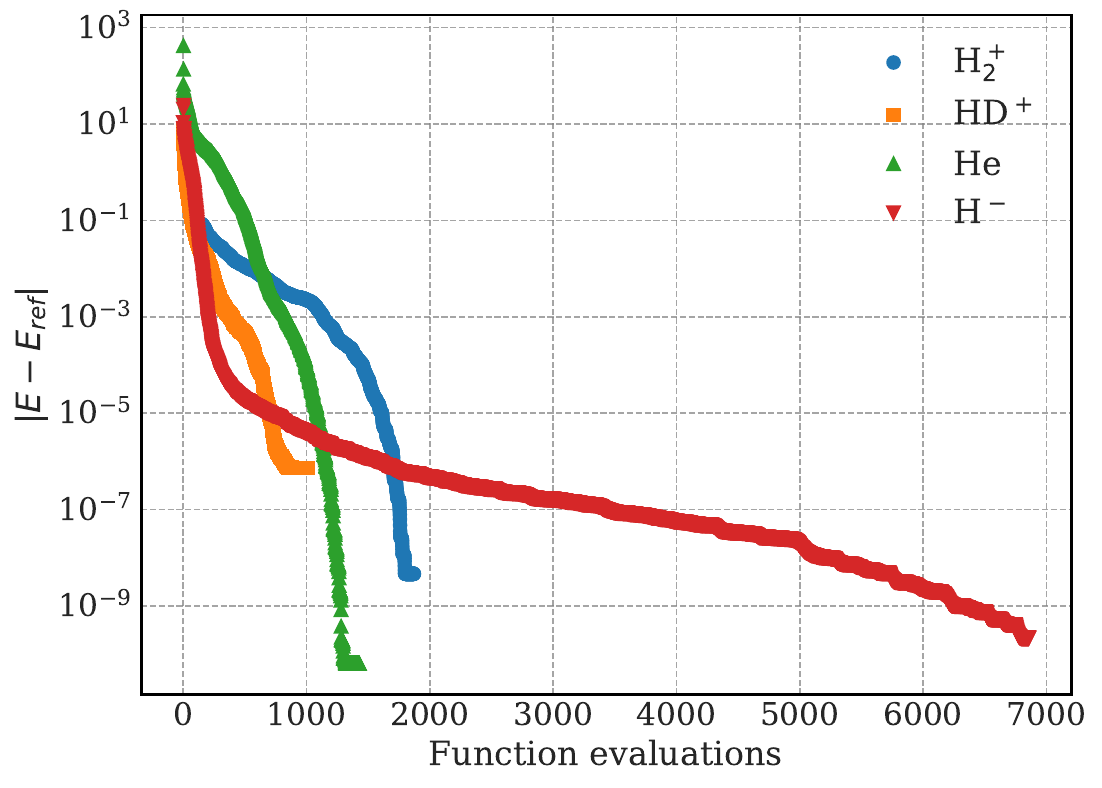}
        \caption{Convergence of NI-DUCC energy error}
        \label{fig:subfig-c}
    \end{subfigure}
    \hfill
    \begin{subfigure}[b]{0.495\textwidth}
        \centering
        \includegraphics[width=\textwidth]{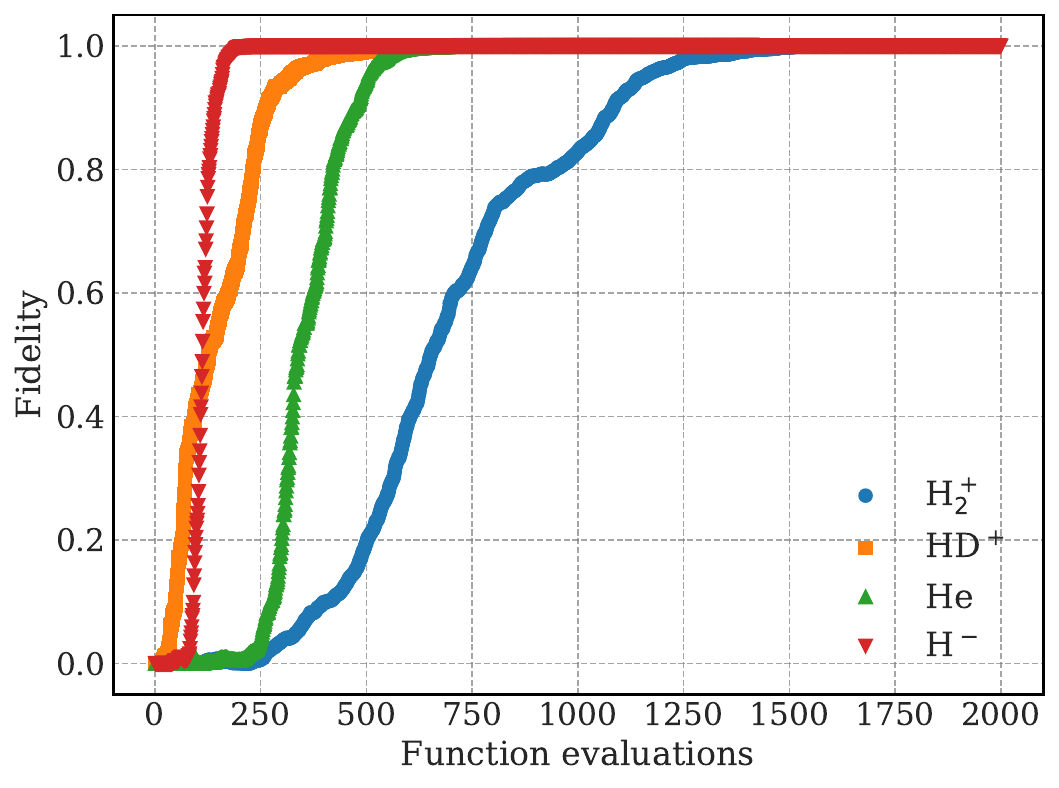}
        \caption{Fidelity of the NI-DUCC wavefunction}
        \label{fig:subfig-d}
    \end{subfigure}
    \caption{(a) Table comparing energies computed using classical computational methods with high-precision reference values. (b) Energy convergence plots for the ground states of H$_2^+$, HD$^+$, He, and H$^-$ obtained via the NI-DUCC-VQE algorithm, using $n=7$ qubits. The energy errors are computed as the difference between the NI-DUCC-VQE result and the reference energy given in (a). (c)
    Convergence of the fidelity, calculated as the overlap $\langle \Psi_j(\vec{\theta}^*) | \Psi_g \rangle$ between the NI-DUCC-VQE state at each optimization step $j$ and the exact ground-state eigenvector $|\Psi_g\rangle$ of the Hamiltonian $H$.}
    \label{fig:full-figure}
\end{figure}

In this section, we analyze the convergence of NI-DUCC-VQE simulations for the ground states of four quantum systems: H$_2^+$, HD$^+$, H$^-$, and He. Each system is represented using 7 qubits, corresponding to 128 basis functions. 
To evaluate the accuracy of NI-DUCC-VQE, we compute the energy error, shown in Fig.~\ref{fig:subfig-c}, as the difference between the NI-DUCC-VQE results and  ultra-accurate reference energies calculated (using classical computing methods) with a much larger basis size. These reference values are listed in Fig.~\ref{fig:subfig-b}.
 The convergence of the NI-DUCC-VQE energies as a function of the number of optimization steps is shown in Fig.~\ref{fig:subfig-c}. Converged values agree to all digits with the results of classical variational computations with $N = 128$, reported in the first row  in Fig.~\ref{fig:subfig-b}. Our NI-DUCC-VQE simulations thus fully match the high precision of classical variational computations, reaching energy errors in the range of $10^{-9}$ a.u. for H$_2^+$, $10^{-7}$ a.u. for HD$^+$, $10^{-10}$ a.u. for H$^-$, and $10^{-11}$ a.u.\ for He. The lower precision achieved in HD$^+$ is linked to the loss of the nuclear echange symmetry.
Moreover, the NI-DUCC-VQE algorithm, relying on the BFGS optimizer for parameter optimization, exhibits highly efficient convergence. For H$_2^+$, HD$^+$, and He, fewer than 2,000 function evaluations were sufficient to reach convergence, while H$^-$ required approximately 7,000 evaluations. The rapid convergence highlights the strength of NI-DUCC’s excitation selection. 
It employs an MCP in which the excitation operators form a set closed under commutation, that ensures a closed Lie-algebraic structure 
(see Section~2 of Ref.~\cite{haidar2025non}), which allows the ansatz to capture 
the essential components of the NI-DUCC wavefunction with high precision 
while avoiding convergence to local minima.
A similar convergence pattern was reported for larger strongly correlated systems such as H$_6$, LiH, and BeH$_2$ (see Fig. 2 in Reference~\cite{haidar2025non}), where the NI-DUCC-VQE with 8 layers was shown to achieve full convergence with only approximately 800 function evaluations. That study attributed the rapid convergence and avoidance of local minima to the use of excitation operators that satisfy a closed Lie-algebraic structure. These operators help eliminate the ordering issues commonly found in many ansätze for strongly correlated systems~\cite{grimsley2019trotterized}. 

Additionally, Fig.~\ref{fig:subfig-d} shows the state fidelity, defined as the overlap between the optimized NI-DUCC-VQE state and the exact ground state eigenvector of the finite-basis Hamiltonian $H$, plotted against the number of function evaluations. Fidelity reaches 1.0 with accuracy only limited by double-precision arithmetic in all four systems, which confirms the robustness and accuracy of the NI-DUCC ansatz.

\begin{figure}[!htbp]
    \centering
    \begin{subfigure}[b]{0.49\textwidth}
        \centering
        \includegraphics[width=\textwidth]{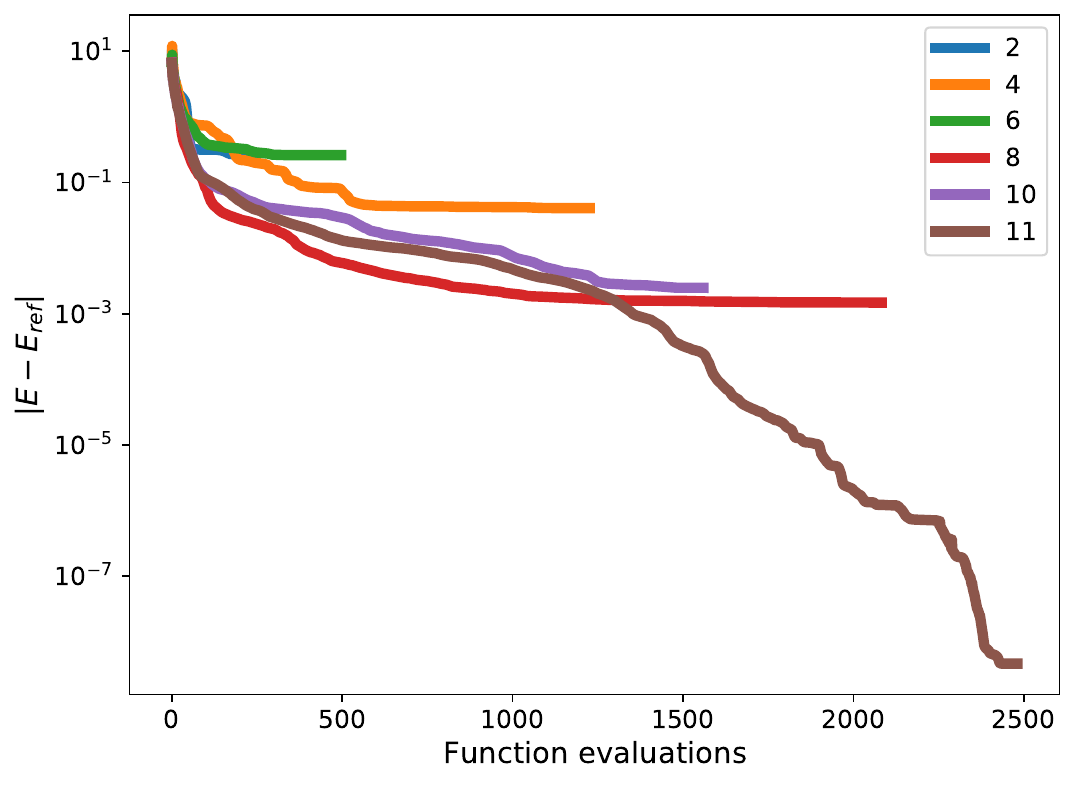}
        \caption{Energy convergence of H$_2^+$ at $n=7$}
        \label{fig:left1}
    \end{subfigure}
     \begin{subfigure}[b]{0.48\textwidth}
        \centering
        \includegraphics[width=\textwidth]{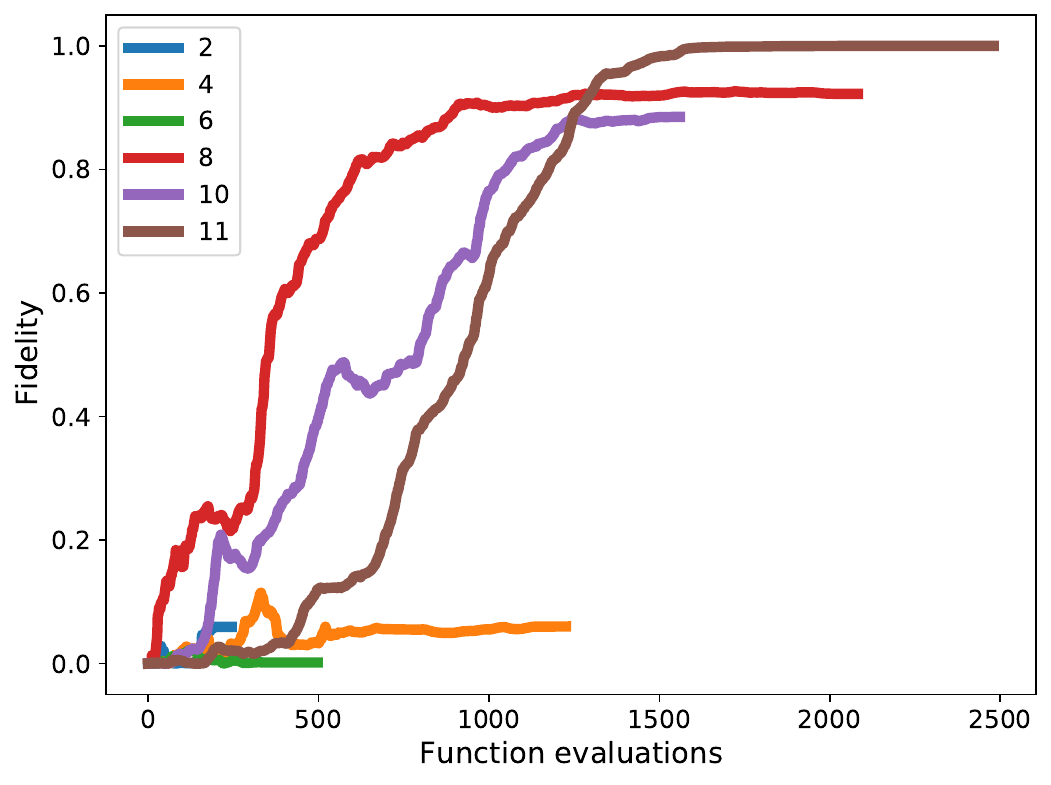}
        \caption{Fidelity of H$_2^+$ wavefunction at $n=7$}
        \label{fig:right1}
    \end{subfigure}
    \begin{subfigure}[b]{0.50\textwidth}
        \centering
        \includegraphics[width=\textwidth]{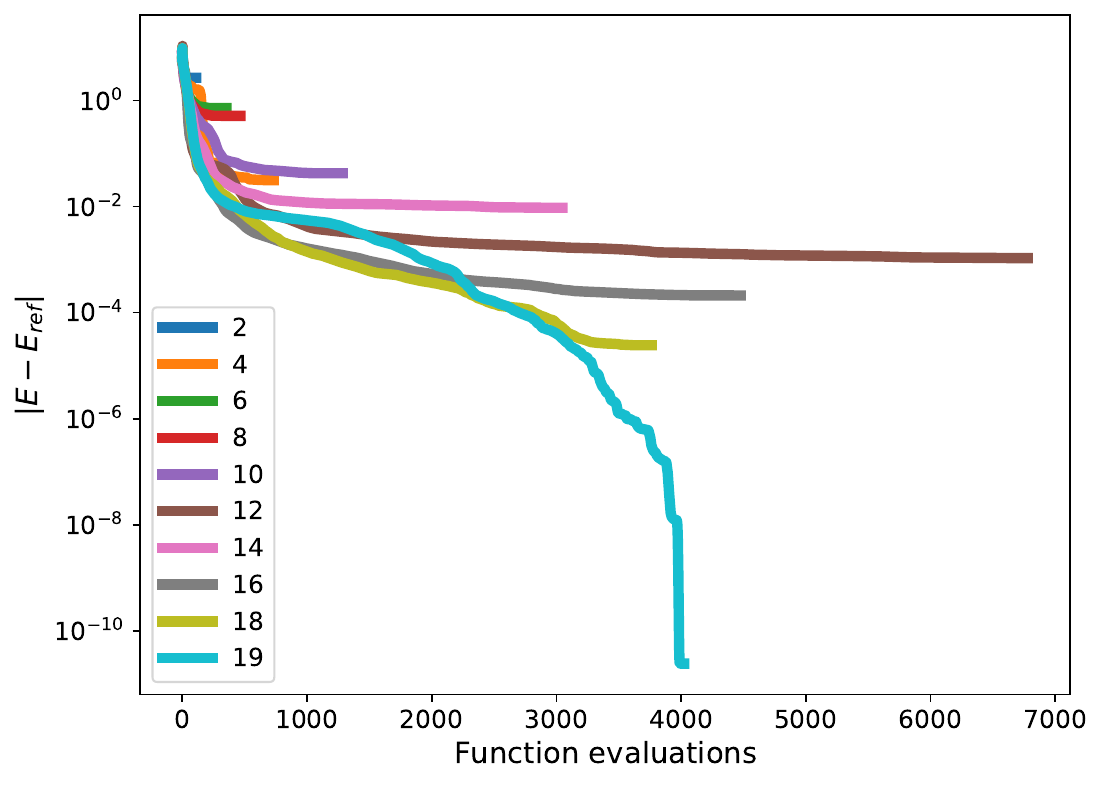}
        \caption{Energy convergence of H$_2^+$ at $n=8$}
        \label{fig:left2}
    \end{subfigure}
     \begin{subfigure}[b]{0.49\textwidth}
        \centering
        \includegraphics[width=\textwidth]{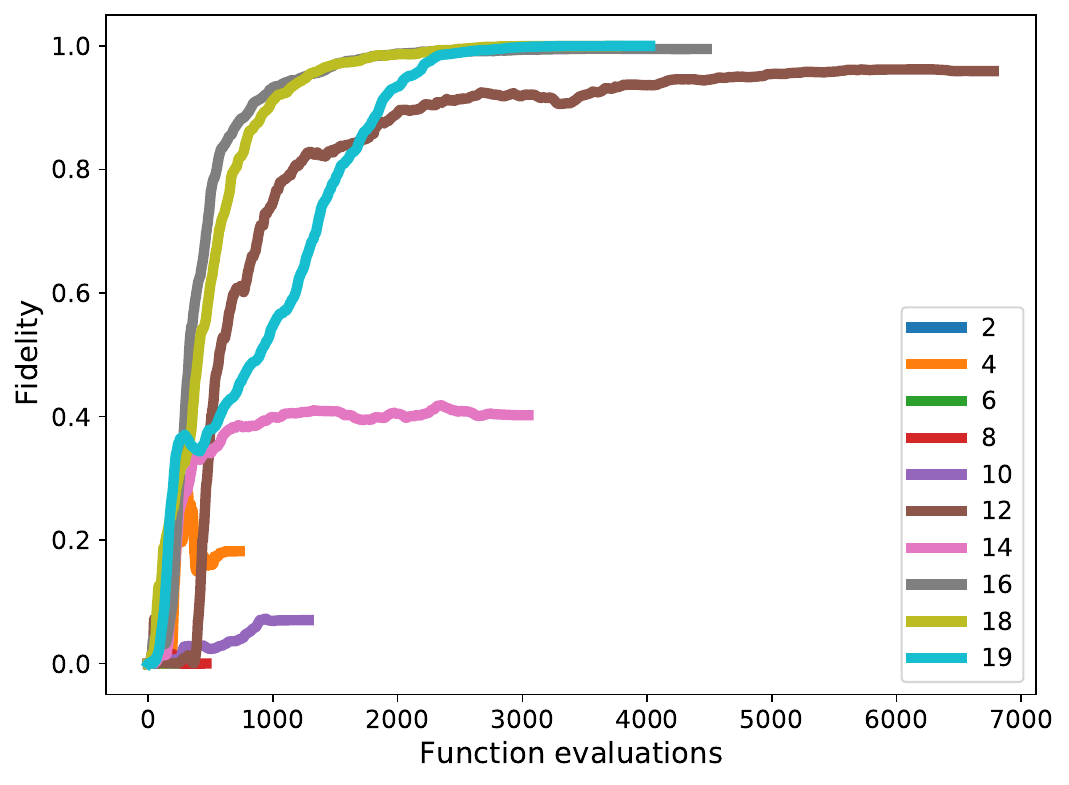}
        \caption{Fidelity of H$_2^+$ wavefunction at $n=8$}
        \label{fig:right2}
    \end{subfigure}    
    \caption{Convergence plots for H$_2^+$ using NI-DUCC-VQE with 7 qubits ($N = 128$) and 8 qubits ($N = 256$) are shown in (a) and (c) for different numbers of layers $k$. The fidelity is displayed in (b) and (d).}
    \label{fig:main}
\end{figure}

\newpage
We also investigated the role of the multi-layered MCP protocol in the construction of the NI-DUCC wavefunction, focusing on the effect of increasing the number of layers $k$. We evaluate this behavior on the H$_2^+$ molecule using $n=7$ qubits ($N=128$ basis set)(Figs.~\ref{fig:left1}-\ref{fig:right1}) and $n=8$ qubits ($N=256$ basis set ) (Figs.~\ref{fig:left2}-\ref{fig:right2}).

For $n=7$, as can be seen in Fig.~\ref{fig:left1}, for $k \leq 10$ the NI-DUCC energy converges only up to a numerical precision in the $10^{-3}$~ a.u. range. However, for $k = 11$, it reaches a much higher precision of a few $10^{-9}$~a.u., matching that of the classical computation.
 The NI-DUCC-VQE energy with $k = 11$ perfectly agrees with the solution from classical diagonalization of the Hamiltonian with $N=128$ (Table~1 in Appendix~\ref{referencesclassical}), with an error below $10^{-14}$ a. u. The calculation uses about 2000 function evaluations and shows no significant plateaus. The fidelity (Fig.~\ref{fig:right1}) improves with increasing $k$. In particular, the $k = 11$ case achieves a fidelity consistent with unity in fewer than 2500 evaluations, considerably outperforming the $k = 8$ and $k = 10$ cases.

In the 8-qubit case, Fig.~\ref{fig:left2} shows that for $k = 19$ the NI-DUCC-VQE energy again accurately matches the classical value (Table 2 in Appendix~\ref{referencesclassical}) and improves precision by nearly two orders of magnitude to a few $10^{-11}$ a.u., compared to $k = 11$ with 7 qubits. The fidelity (Fig.~\ref{fig:right2}) reaches a value consistent with unity after 4000 evaluations for $k=19$, largely outperforming lower $k$ values.

The difference in convergence performance across layers becomes relevant if the computation is terminated before full convergence. For some applications, reaching moderate numerical precision may be sufficient, making lower $k$ values acceptable. However, for tasks that require high-precision results, increasing the number of excitation layers becomes advantageous. In what follows, we adopt the NI-DUCC wavefunction with $k = 11$ in the 7-qubit case (or $k=19$ if the 8-qubit case is considered).
 
\subsection{NI-DUCC-VQE versus Qubit-ADAPT-VQE}

Since resource efficiency is crucial for assessing the suitability of an ansatz on current NISQ devices, this section compares the performance of NI-DUCC-VQE with the Qubit-ADAPT-VQE variational ansatz~\cite{tang2021qubit}. The comparison focuses on key metrics—number of parameters, CNOT gate counts, and function evaluations required for convergence. The lower the resource requirements, the more compatible the ansatz is with NISQ-era hardware. As we will see in our analysis, NI-DUCC-VQE outperforms Qubit-ADAPT-VQE, particularly requiring significantly fewer function evaluations.
 NI-DUCC employs a fixed ansatz structure based on a Lie-algebraic disentangled unitary expansion (see Section~\ref{sec:NI-DUCC} for details). In contrast, Qubit-ADAPT-VQE constructs the ansatz adaptively by selecting operators that yield the largest energy gradient. It can produce compact, problem-specific circuits, but it imposes a high measurement cost due to the gradient evaluations over the entire operators pool at each iteration. We first outline the Qubit-ADAPT-VQE framework, and then present a direct comparison of the resource requirements for both algorithms.
The Qubit-ADAPT-VQE method builds the variational ansatz iteratively by selecting one 
operator at a time from a predefined pool. In this work, we use the MCP~\cite{tang2021qubit} 
for both NI-DUCC and Qubit-ADAPT-VQE to ensure a fair comparison. At each iteration, 
Qubit-ADAPT-VQE evaluates the energy gradient associated with every Pauli operator 
\(\hat{A}_j\) in the MCP, which can be expressed as
\[
g_j = \left|\left\langle \psi(\boldsymbol{\theta}) \middle| \left[ \hat{H}, \hat{A}_j \right] \middle| \psi(\boldsymbol{\theta}) \right\rangle\right|,
\]
and then selects the operator with the largest gradient magnitude to extend the ansatz.  The gradient norm is denoted as $g(\epsilon)$ and defined by:
\begin{equation}
g = \| \vec{g} \| = \sqrt{\sum_j |g_j|^2}
\label{eq:gradient_norm}
\end{equation}
 The algorithm continues to grow the ansatz until $g$ falls below a user-defined threshold $\epsilon$. In contrast, NI-DUCC does not rely on gradient evaluations; instead, it systematically 
generates its excitation operators from the Lie-algebraic structure of the MCP, thereby 
circumventing the costly gradient computations required in ADAPT-VQE.

\begin{figure}[!h] 
    \centering
    \begin{subfigure}[t]{0.48\textwidth}
        \centering
        \includegraphics[width=\linewidth]{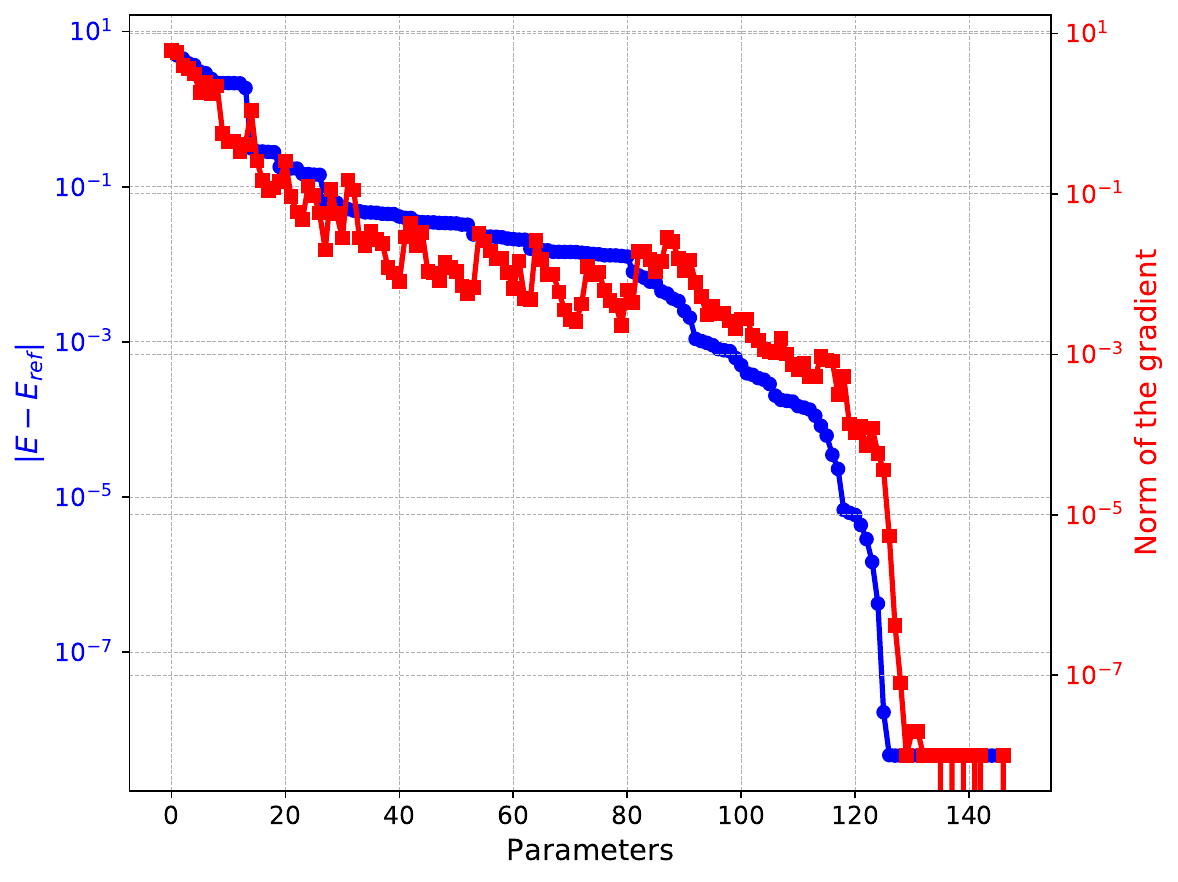}
        \caption{Convergence of the energy (blue curve, left axis) and gradient norm (red curve, right axis) as a function of the number of parameters.}
        \label{fig:qubit_adapt}
    \end{subfigure}
    \hfill
    \begin{subfigure}[t]{0.49\textwidth}
        \centering
        \includegraphics[width=\linewidth]{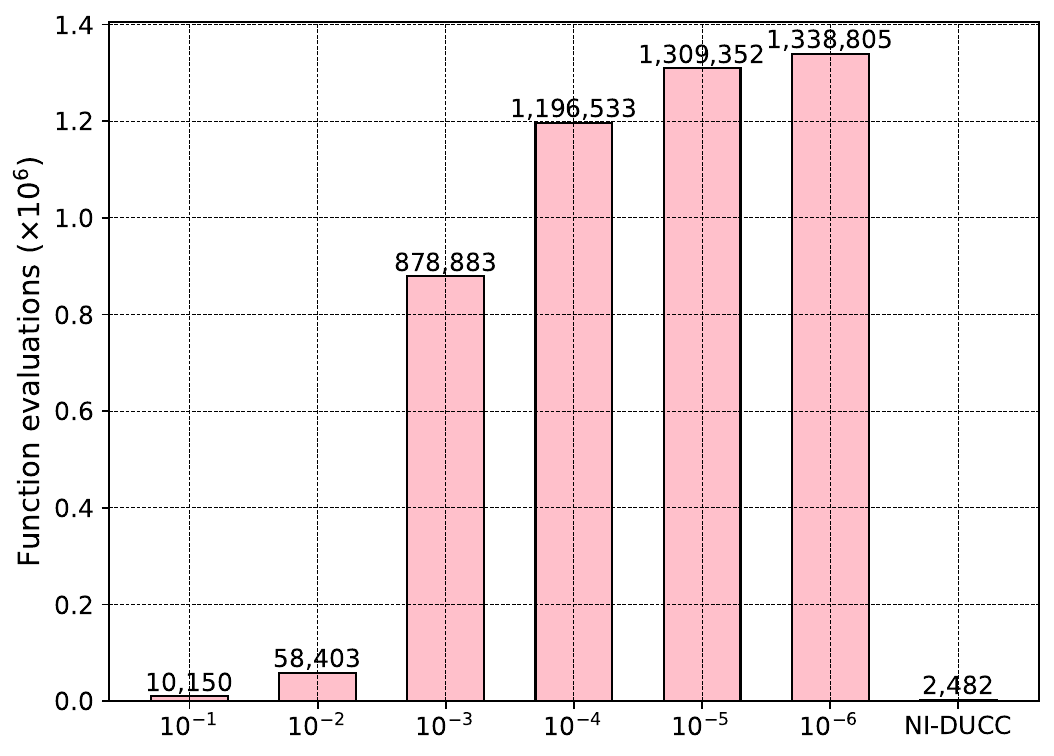}
        \caption{Comparison of function evaluation counts between Qubit-ADAPT and NI-DUCC Methods. For ADAPT-VQE, the horizontal axis indicates the value of the convergence threshold $\epsilon$.}
        \label{fig:histogram}
    \end{subfigure}

    \caption{Convergence of Qubit-ADAPT-VQE for the 
H\textsubscript{2}\textsuperscript{+} molecule (7 qubits, $N=128$ basis functions), and resource comparison with NI-DUCC-VQE.
Panel (a) shows the energy error (in a.u.) of Qubit-ADAPT-VQE relative to the 
reference value, together with the corresponding gradient norm $g$ (right axis) 
as the ansatz is iteratively expanded. In Qubit-ADAPT-VQE, one operator is added 
per iteration; the number of parameters, which is equal to the number of selected operators, thus corresponds to the number of iterations. 
In panel (b), the first six columns show the total number of required function evaluations in Qubit-ADAPT for the gradient norm $g$ to reach the indicated 
threshold values $\epsilon = 10^{-1}, 10^{-2}, \ldots$. The last column shows the number of function evaluations in NI-DUCC-VQE with $k=12$ layers. 
Function evaluations correspond to the optimization steps performed using the BFGS optimizer.}
    \label{fig:main1}
\end{figure}

To compare performances, we use  the H$_2^+$ system encoded with 7 qubits. Fig.~\ref{fig:qubit_adapt} reports Qubit-ADAPT-VQE energy errors relative to the reference value and the corresponding gradient norm. Energy precision improves as the gradient norm decreases. In terms of parameter count, Fig.~\ref{fig:qubit_adapt} shows that Qubit-ADAPT-VQE requires 144 iterations, which corresponds to 144 variational parameters, to reach the same precision (a few $10^{-9}$ a.u.) as NI-DUCC-VQE with $k = 12$ (Fig.~\ref{fig:left1}).
Fig.~\ref{fig:histogram} shows the total number of function evaluations across various $\epsilon$ values, benchmarked against NI-DUCC-VQE with $k = 11$. Qubit-ADAPT-VQE requires 10,151 evaluations with BFGS to reach a threshold value $\epsilon = 0.1$ a.u., corresponding to a moderate precision of $10^{-1}$ a.u. on the energy (compare the blue and red curves in Fig.~\ref{fig:qubit_adapt}), while NI-DUCC-VQE with $k = 11$ achieves high precision  with only 2,842 evaluations. Qubit-ADAPT would needs over $10^6$ evaluations for similar precision. This demonstrates the superior efficiency of NI-DUCC over Qubit-ADAPT-VQE in achieving rapid convergence with high precision and without any gradient measurement overhead.
The performance of the NI-DUCC approach has already been demonstrated in prior work by Haidar et al.~\cite{haidar2025non}, where NI-DUCC outperforms Qubit-ADAPT-VQE in symmetry-constrained molecular systems such as H$_6$ and BeH$_2$ (see Fig. 4 in that reference). These results highlight NI-DUCC’s potential as a systematic method for reaching high precision within three-body systems and beyond.

Furthermore, from a hardware implementation perspective, resource demands go beyond parameter count and function evaluations. On Noisy Intermediate-Scale Quantum (NISQ) devices~\cite{google_quantum_ai_datasheet}, the number of two-qubit entangling gates—particularly CNOT gates—is a key factor for the circuit's fidelity, due to their susceptibility to noise and their relatively higher error rates compared to single-qubit gates.
To estimate the CNOT gate cost, we follow Refs.~\cite{mcardle2020quantum,whitfield2011simulation,hempel2018quantum}. The CNOT cost for each NI-DUCC Pauli excitation in the MCP, with length $p$ and an odd number of $Y$ terms, scales as $2p - 2$. In our study of the H$_2^+$ system, both algorithms require 132 Pauli excitations to reach optimal precision.  Therefore, the total CNOT cost for each method is approximately $(2p - 2) \times 132$. This demonstrates that NI-DUCC achieves equivalently efficient CNOT scaling without the iterative and measurement-intensive procedures that Qubit-ADAPT-VQE requires.

\section{Evaluation of delta-function operators within the NI-DUCC-VQE framework}

The nonrelativistic energies obtained from the three-body Hamiltonian in Eq.~(\ref{Hamiltoniantotal}) can be improved by computing relativistic and QED corrections (see e.g.~\cite{Korobov2006,Korobov2021,haidar2022higher,Haidar2021} for the case of HMI), many of which can be expressed, in a perturbative framework, in the form of delta-function  expectation values, $\langle \delta(\mathbf{r}_a) \rangle$. In this section, we evaluate these terms using the NI-DUCC-VQE algorithm to investigate its numerical precision and stability.

We begin by computing the matrix elements
\begin{equation}
D_{ij} = \langle \phi'_i | \delta(\mathbf{r}_a) | \phi'_j \rangle, \quad a = 1, 2,
\end{equation}
where $\phi'_i$ denote the orthonormalized basis functions, obtained through canonical orthonormalization of the basis defined in Eq.~(\ref{basisfunc}) (see Appendix~\ref{app_co}). The resulting matrix $D$ is symmetric: $D_{ij} = D_{ji}$, then the expectation value of the delta operator over the variational wavefunction (\ref{Totalwavefunction1}) is given by
\begin{equation} \label{eq:detla-exp}
\langle \phi | \delta(\mathbf{r}_a) | \phi \rangle = \sum_{i,j=1}^{N} c'_i c'_j D_{ij} = 2 \sum_{i>j} c'_i c'_j D_{ij} + \sum_{i} {c'_i}^2 D_{ii}.
\end{equation}
 Numerical values of these delta-function expectation values, obtained by classical computational methods, are presented in Table~\ref{tab-delta} in Appendix~\ref{referencesclassical}.

\begin{figure}[htbp]
    \centering
    \includegraphics[width=0.5\textwidth]{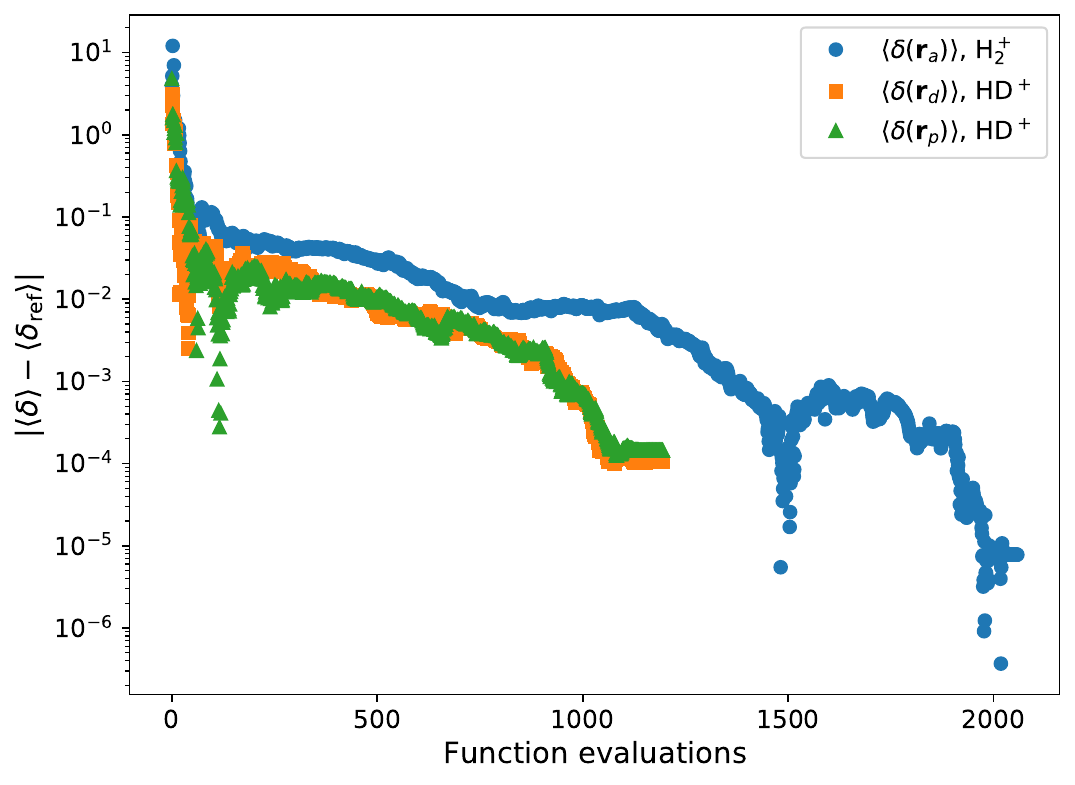} 
     \caption{\RaggedRight Convergence of the delta-function expectation values $\langle \delta(\mathbf{r}_a) \rangle$ is shown for the ground states of H$_2^+$ and HD$^+$. The calculations are done with 7 qubits ($N=128$) and 12 NI-DUCC-VQE layers. The errors (vertical axis) are computed as the difference between the NI-DUCC-VQE result and the reference values given in Table~11 of Appendix~\ref{referencesclassical}.}
    \label{fig:delta_errors}
\end{figure}

To calculate these expectation values using the NI-DUCC-VQE method, we map the delta operators into a qubit-compatible form using binary encoding, as described in Sec.~\ref{first_quantizedsec} (see Eq.~(\ref{first_quantized}) and following text), where it was applied to the non-relativistic Hamiltonian. This step expresses the delta-function terms as sums of Pauli operators and completes the VQE procedure. To test this, 
we apply NI-DUCC-VQE to the ground states of H$_2^+$ and HD$^+$. As shown in Fig.~\ref{fig:delta_errors}, the algorithm achieves a precision of $10^{-6}$~a.u. after approximately 2000 function evaluations for H$_2^+$. For HD$^+$, convergence is reached in around 1200 evaluations for both $\delta(\mathbf{r}_p)$ and $\delta(\mathbf{r}_d)$. 
These results demonstrate that NI-DUCC-VQE can accurately compute  perturbative corrections to the energy of three-body systems, going beyond the Born--Oppenheimer approximation in the molecular systems studied here. Extension of this work to other effective operators appearing in energy corrections as well as fine and hyperfine structures could thus be a promising direction for future investigation.

\newpage
\section{Conclusion}
In this work, we applied NI-DUCC-VQE to simulate four benchmark quantum systems—H$_2^+$, HD$^+$, H$^-$, and He, going beyond the Born–Oppenheimer approximation in the case of molecular ions, using a first-quantized Hamiltonian. The method achieved high precision, with energy errors in the range of of $10^{-9}$ a.u. for H$_2^+$, $10^{-7}$ for HD$^+$, $10^{-10}$ for H$^-$, and $10^{-11}$ for He. State fidelities reached 1.0 with accuracy only limited by double-precision
arithmetic across all systems.  These results validate the accuracy and robustness of the NI-DUCC ansatz. The algorithm requires only a few thousand function evaluations, while Qubit-ADAPT-VQE needs over $10^6$ for similar precision. This efficiency comes from the synergy between the linearly scaling MCP of  excitations and binary encoding of the first-quantized Hamiltonian. The qubit count scales as $\log_2 N$, supporting larger basis sets with shallow circuits. The Lie-algebraic closure ensures compact, gradient-free ansatz construction and avoids barren plateaus. Convergence remains stable as basis size increases, which is a promising sign for scaling to more complex systems beyond three-body cases. Moreover, NI-DUCC benchmarks on perturbative corrections, here performed in the case of  delta-function operators, show that they can be obtained with high precision, illustrating  the potential of our method to compute energy corrections beyond the nonrelativistic (Schr\"odinger) level.

A natural next step is to apply NI-DUCC-VQE to four-body systems, such as the H$_2$ molecule and three-electron atoms. In Ref.~\cite{Volkmann2024}, promising results for H$_2$ molecule were reported using Qubit-ADAPT-VQE in conjunction with explicitly correlated basis functions (EC) in a first-quantized framework. However, that work was limited by the Born--Oppenheimer approximation and suffered from bottlenecks in gradient evaluation because of the use of Qubit-ADAPT-VQE.
In contrast, as demonstrated in our work for H$_2^+$ and HD$^+$, NI-DUCC-VQE avoids these bottlenecks and offers a viable route toward treating the H$_2$ molecule and potentially, more complex systems beyond the Born-Oppenheimer approximation using a first-quantized representation.


%

\newpage

\appendix

\section{Canonical orthogonalization} \label{app_co}

We desire to convert a generalized eigenvalue problem
\begin{equation}\label{co_eq1}
    Ax=aBx,    
\end{equation}
where $B$ is the overlap matrix, into a conventional eigenvalue problem of the form $A^{\prime}x^{\prime}=ax^{\prime}$, where the eigenvalue $a$ is unchanged. For this purpose we need a matrix $U$ such that
\begin{equation}\label{co_eq2}
    U^TBU=I    
\end{equation}
According to the canonical orthogonalization method proposed by Lowdin (1993)\cite{lowdin1993some}, $U$ is given by
\begin{equation}\label{co_eq3}
    U=\Delta D,
\end{equation}
where $\Delta$ is a diagonal matrix that contains the inverse of the square root of the eigenvalues of $B$, the columns of $D$ contain the eigenvectors of $B$, and $U^{T}U=1$. We multiply Eq.~(\ref{co_eq1}) by $U^{T}$ on the left and insert the identity to get
\begin{equation}\label{co_eq4}
    U^TAUU^Tx=aU^TBUU^Tx.    
\end{equation}
Finally, we use Eq.~(\ref{co_eq2}) in Eq.~(\ref{co_eq4}) to get
\begin{equation}\label{co_eq5}
    A^{\prime}x^{\prime}=ax^{\prime},    
\end{equation}
where $A^{\prime}=U^TAU$, and $x^{\prime}=U^Tx$. The generalized eigenvalue problem of Eq.~(\ref{co_eq1}) is thus converted to a conventional eigenvalue problem.

\section{Reference classical computation results for three-body systems}

We give in this Appendix the optimized basis sets used to construct the Hamiltonian and overlap matrices [see Eq.~(\ref{eq-generalized eig})], from which the transformed Hamiltonian used for the quantum computations [Eq.~(\ref{first_quantized})] is obtained by applying the orthonormalization procedure described in Appendix~\ref{app_co}. The Tables are constructed as follows: each line describes one subset of the total basis set, where columns $1-6$ contain the optimized intervals for the real and imaginary parts of the exponents $\alpha$, $\beta$, $\gamma$, and the last column gives the number of basis functions in the subset. The last line of the Tables gives the energy value obtained with this basis set using the classical variational method introduced in~\cite{Korobov2000}, together with the energy error, which is determined by comparing with an ultra-accurate reference value obtained with a much larger basis set.\\
Finally, the last Table~\ref{tab-delta} contains delta-function expectation values obtained by classical computations with $N=128$ and $N=256$ basis sets, together with highly precise reference values from the literature.

\label{referencesclassical}
\begin{table}[h!]
\footnotesize
\begin{tabular}{ccccccc}
\hline
${\rm Re} (\alpha)$  & ${\rm Im} (\alpha)$   & ${\rm Re} (\beta)$   & ${\rm Im} (\beta)$    & ${\rm Re} (\gamma)$  & ${\rm Im} (\gamma)$   & $n$  \\
$[1.07654, 1.78066]$ & $[-0.01940, 0.05276]$ & $[0.07667, 0.88031]$ & $[-0.00034, 0.01760]$ & $[2.65129, 3.06464]$ & $[0.52900, 7.16818]$  & $56$ \\
$[0.95374, 1.07476]$ & $[0.00000, 0.03316]$  & $[0.28271, 0.46185]$ & $[-0.00357, 0.00218]$ & $[2.97254, 3.65661]$ & $[2.49296, 11.55876]$ & $36$ \\
$[1.30899, 1.35858]$ & $[-0.01408, 0.02912]$ & $[0.11332, 0.46886]$ & $[0.01506, 0.02934]$  & $[2.85000, 3.09541]$ & $[1.19531, 9.81965]$  & $22$ \\
$[1.18347, 1.25594]$ & $[-0.05596, 0.00581]$ & $[0.21488, 0.29078]$ & $[-0.00677, 0.00112]$ & $[2.53124, 2.84387]$ & $[0.68702, 2.29223]$  & $12$ \\
$[0.81427, 1.73937]$ & $[0.00000, 0.07147]$  & $[0.04726, 0.29998]$ & $[-0.01306, 0.00566]$ & $[2.02013, 4.64661]$ & $[1.11470, 1.11641]$  & $2$ \\
\hline \noalign{\vskip 1mm}
Energy (a.u.)           & $-0.597 \, 139 \, 058 \, 413$    & \multicolumn{2}{c}{Error on energy (a.u.)} & $4.7 \times 10^{-9}$ & \\[1mm]
\hline
\end{tabular}
\caption{Optimized basis set for the ground state of H$_2^+$ with 7 qbits ($N = 128$). The CODATA 2018 value of the proton-electron mass ratio is used. The exact ground-state energy is $E = -0.597 \, 139 \, 063 \, 080 \ldots$ \label{tab-h2plus-v0-N128}}
\end{table}
\begin{table}[h!]
\footnotesize
\begin{tabular}{ccccccc}
\hline
${\rm Re} (\alpha)$  & ${\rm Im} (\alpha)$   & ${\rm Re} (\beta)$   & ${\rm Im} (\beta)$    & ${\rm Re} (\gamma)$  & ${\rm Im} (\gamma)$   & $n$  \\
$[1.19075, 1.65638]$ & $[-0.02595, 0.03948]$ & $[0.03516, 1.33377]$ & $[-0.00078, 0.01667]$ & $[2.98179, 3.28156]$ & $[0.02937, 7.39889]$  & $112$ \\
$[1.02251, 1.03871]$ & $[0.00000, 0.00381]$  & $[0.31684, 0.38798]$ & $[-0.00389, 0.00222]$ & $[3.65150, 3.68554]$ & $[2.68553, 11.57492]$ & $72$ \\
$[1.35418, 1.43248]$ & $[-0.01929, 0.02702]$ & $[0.15301, 0.47203]$ & $[0.01553, 0.03148]$  & $[2.75186, 3.71197]$ & $[0.47651, 12.16416]$ & $44$ \\
$[1.13639, 1.14046]$ & $[-0.05515, 0.00557]$ & $[0.20079, 0.25503]$ & $[-0.00662, 0.00092]$ & $[3.38954, 3.51040]$ & $[0.87464, 2.76446]$  & $24$ \\
$[0.79078, 1.62120]$ & $[0.03811, 0.07337]$  & $[0.14785, 0.24217]$ & $[-0.01397, 0.00599]$ & $[2.60898, 5.49922]$ & $[1.16318, 1.17671]$  & $4$ \\
\hline \noalign{\vskip 1mm}
Energy (a.u.)           & $-0.597 \, 139 \, 063 \, 056$    & \multicolumn{2}{c}{Error on energy (a.u.)} & $2.4 \times 10^{-11}$ & \\[1mm]
\hline
\end{tabular}
\caption{Same as Table~\ref{tab-h2plus-v0-N128}, with 8 qbits ($N = 256$). \label{tab-h2plus-v0-N256}}
\end{table}
\vspace{4mm}
\begin{table}
\footnotesize
\begin{tabular}{ccccccc}
\hline
${\rm Re} (\alpha)$  & ${\rm Im} (\alpha)$   & ${\rm Re} (\beta)$   & ${\rm Im} (\beta)$    & ${\rm Re} (\gamma)$  & ${\rm Im} (\gamma)$   & $n$  \\
$[0.99855, 1.36647]$ & $[-0.05423, 0.06681]$ & $[0.33903, 0.41092]$ & $[-0.02864, 0.02492]$ & $[2.94950, 3.05629]$ & $[0.88001, 8.74844]$  & $44$ \\
$[0.13197, 0.90499]$ & $[-0.02043, 0.00050]$ & $[0.82133, 1.38484]$ & $[0.00000, 0.00000]$  & $[2.62231, 2.84177]$ & $[2.13254, 10.78452]$ & $44$ \\
$[1.14875, 1.15717]$ & $[-0.00035, 0.00984]$ & $[0.33390, 0.36618]$ & $[-0.01329, 0.01905]$ & $[2.75207, 3.49929]$ & $[0.83619, 4.08752]$  & $20$ \\
$[0.00000, 0.46873]$ & $[-0.00765, 0.00112]$ & $[1.08927, 1.44436]$ & $[-0.03147, 0.01546]$ & $[2.63231, 3.16904]$ & $[1.03728, 4.64237]$  & $20$ \\
\hline \noalign{\vskip 1mm}
Energy (a.u.)           & $-0.597 \, 897 \, 235 \, 547$    & \multicolumn{2}{c}{Error on energy (a.u.)} & $7.3 \times 10^{-7}$ & \\[1mm]
\hline
\end{tabular}
\caption{Optimized basis set for the ground state of HD$^+$ with 7 qbits ($N = 128$). The CODATA 2018 values of the deuteron-electron and proton-electron mass ratio are used. The exact ground-state energy is $E = -0.597 \, 897 \, 968 \, 103 \ldots$ \label{tab-hdplus-v0-N128}}
\end{table}

\begin{table}
\footnotesize
\begin{tabular}{ccccccc}
\hline
${\rm Re} (\alpha)$   & ${\rm Im} (\alpha)$   & ${\rm Re} (\beta)$    & ${\rm Im} (\beta)$    & ${\rm Re} (\gamma)$  & ${\rm Im} (\gamma)$   & $n$  \\
$[1.74188, 2.25322]$  & $[0.18652, 0.46520]$  & $[1.28024, 2.53851]$  & $[-0.02126, 0.37233]$ & $[0.11632, 0.21375]$ & $[0.01845, 0.30082]$  & $52$ \\
$[3.00721, 5.11414]$  & $[-0.00008, 0.00098]$ & $[1.81077, 3.05217]$  & $[-0.23520, 0.63307]$ & $[0.19661, 0.79442]$ & $[-0.01931, 0.86401]$ & $42$ \\
$[0.05586, 19.36455]$ & $[0.15767, 0.48686]$  & $[2.58347, 13.02001]$ & $[0.00000, 0.31394]$  & $[1.21826, 4.29668]$ & $[-0.38445, 1.82136]$ & $34$ \\
\hline \noalign{\vskip 1mm}
Energy (a.u.)           & $-2.903 \, 724 \, 376 \, 970$    & \multicolumn{2}{c}{Error on energy (a.u.)} & $6.4 \times 10^{-11}$ & \\[1mm]
\hline
\end{tabular}
\caption{Optimized basis set for the ground ($1S^1$) state of He with infinite nuclear mass, using 7 qbits ($N = 128$). The exact energy is $E = -2.903 \, 724 \, 377 \, 034 \, 119 \ldots$ \label{tab-he-N128}}
\end{table}

\begin{table}
\begin{tabular}{ccccccc}
\hline
${\rm Re} (\alpha)$   & ${\rm Im} (\alpha)$   & ${\rm Re} (\beta)$     & ${\rm Im} (\beta)$    & ${\rm Re} (\gamma)$  & ${\rm Im} (\gamma)$   & $n$  \\
$[0.14510, 1.31809]$  & $[-0.04241, 0.12121]$ & $[1.03474, 1.08192]$   & $[-0.02314, 0.19099]$ & $[0.04439, 0.04621]$ & $[-0.01740, 0.04416]$ & $52$ \\
$[1.35801, 4.29001]$  & $[-0.14619, 0.74523]$ & $[1.05935, 4.45260]$   & $[-0.00718, 0.25993]$ & $[0.05716, 0.96584]$ & $[-0.11928, 0.65945]$ & $42$ \\
$[7.30700, 31.99866]$ & $[0.21476, 0.88766]$  & $[20.81734, 26.21358]$ & $[0.29711, 0.33078]$  & $[0.35002, 3.58686]$ & $[-0.38968, 1.65546]$ & $34$ \\
\hline \noalign{\vskip 1mm}
Energy (a.u.)           & $-0.527 \, 751 \, 016 \, 439$    & \multicolumn{2}{c}{Error on energy (a.u.)} & $1.1 \times 10^{-10}$ & \\[1mm]
\hline
\end{tabular}
\caption{Optimized basis set for the ground state of H$^-$ with infinite nuclear mass, using 7 qbits ($N = 128$). The exact energy is $E = -0.527 \, 751 \, 016 \, 544 \, 377 \ldots$ \label{tab-hminus-N128}}
\end{table}

\begin{table}
\begin{tabular}{cccc}
\hline
$N$      & $\langle \delta(\mathbf{r}_a) \rangle$, H$_2^+$ & $\langle \delta(\mathbf{r}_d) \rangle$, HD$^+$ & $\langle \delta(\mathbf{r}_p) \rangle$, HD$^+$ \\
\hline
128      &  $0.2067442803$                                 &  $0.2074523162$                                &  $0.2071868989$                                \\
ref. &  $0.20673647629$                                &  $0.20734814178$                               &  $0.20704259948$                               \\
\hline
\end{tabular}
\caption{Delta-function expectation values in the molecular ground state obtained by classical variational computations. In HD$^+$, $\mathbf{r}_d$ ($\mathbf{r}_p$) is the position of the electron with respect to the deuteron (proton). Very precise values from~\cite{Aznabayev2019} are reported in the last line.  \label{tab-delta}}
\end{table}

\end{document}